\documentclass[12pt]{article}
\usepackage{graphicx}
\usepackage{multicol}
\usepackage{pdfpages}
\usepackage{titlesec}
\usepackage{braket}
\usepackage{amssymb}
\usepackage{amsmath}
\usepackage{float}
\usepackage{aliascnt}
\usepackage[top=2.54cm,bottom=2.54cm,left=3.175cm,right=2.54cm]{geometry}
\usepackage{times}
\usepackage{setspace}
\usepackage[hidelinks]{hyperref}
\usepackage{tocloft}

\begin{document}

\begin{titlepage}   
\begin{center}
    \vspace{1cm}
    \Large
    \textbf{Showing Ambiguity in the Pilot-Wave Theory Equations of Motion via the Derivation of Unique Scalar Fields Using a 2D Quantum Harmonic Oscillator}\\
    \large
    \vspace{1.25cm}
    \textbf{Connell Bristow}\\connellbristow@gmail.com\\
    \vspace{1cm}19th June 2023\\
\end{center}
\end{titlepage}

\pagenumbering{roman}
\begin{abstract}
\noindent In De Broglie-Bohm Pilot-Wave Theory unique equations of motion and scalar fields for a particle can be formulated. This is done by finding a solution for a divergence free probability density current $\vec{J}(r,t)$ and then dividing by the Born Rule $|\vec{\Psi}(r,t)|^{2}$ for velocity arising from Hamilton-Jacobi mechanics. This addition of divergence free probability current would still satisfy the Schrodinger continuity equation. It was found that for a 2D polar system a divergence free probability density is the gradient rotated by the matrix ($S$) applied to a scalar field as such $S\vec{\nabla}\varphi(r)$. A variety of equations of motion were used such that the dimensions of the equations are equivalent to a velocity, different scalar fields were then derived for a 2D quantum harmonic oscillator. It was found that for each unique velocity there was a unique scalar field associated with it, as such the equation of motion for a particle in De Broglie-Bohm pilot wave theory could be theoretically dependent on any arbitrary number of scalar fields. This ultimately means the equations of motion could be ambiguous as there are multiple mathematically valid solutions.  
\end{abstract}
\newpage
\tableofcontents
\newpage
\pagenumbering{arabic}

\section{Introduction}
\subsection{Historical Context}
\label{Intro}
De Broglie-Bohm pilot wave theory was founded by Louis De-Broglie in the 1920s and discarded and then redeveloped by chance by David Bohm in the 1950s under the name of Bohemian Mechanics as an interpretation of quantum mechanics. This interpretation was met with backlash as it is explicitly non-local which contradicted special relativity, which already had been widely accepted by the scientific community. The interpretation is a hidden variable theory assuming that the universe follows a guiding wave function that is influenced by all particles instantaneously in the universe and knowing the initial conditions before hand would tell you the outcome on a measurement \cite{holland1995quantum}. This essentially painted the quantum world as deterministic which was a direct counter argument to the standard consensus at the time, the Copenhagen Interpretation. In recent times the De Broglie-Bohm pilot wave theory has recently made a resurgence with recent work being done to merge special relativity with pilot-wave theory \cite{nikolic2019relativisticquantummechanicsquantum} \cite{D_rr_1999}. The discovery and experimental evidence of quantum teleportation \cite{EQTNature309}, has provided further proof of non-local effects helping to push Pilot-Wave theory more seriously among physicists.

\subsection{Previous Work and Alterations}
In order for simplicity and some efficient cancellations, the value for $\varphi$ used was the Born rule itself, which produced some very straightforward equations. These solutions were originally found by Kylie Bristow in 2023 as part of their masters thesis. At the time the equations were accidentally misinterpreted as velocity equations when the report was submitted. After doing dimensional analysis on those equations it was found that the equations were not correctly aligned with velocity dimensions \cite{Kylie2023}. This report aims to correct that by doing the same method, and then fixing the equation by multiplying by the dimensions needed to adhere to a velocity, then simply working in reverse to find the scalar field $\varphi$ needed to achieve that equation of motion. Different working scalar fields would then further confirm there is ambiguity in the equations of motion.

\subsection{Objectives and Aims}
This report will take the framework developed by David Bohm and expand upon it, using the original mathematical techniques and logic used to derive the interpretation. Specifically this report will further probe the mathematics of pilot-wave theory, using a 2D quantum harmonic oscillator as the system used to explore the mathematics. The aim of this report is to show how unique scalar fields can be derived from various equations of motion, showing that there is an underlying ambiguity in Pilot-Wave Theory. Providing new insights into the workings of the interpretation and questioning the validity of the theory. When deriving the equation for a 2D divergence free probability density $\vec{J}(r,t)$ it was found that an arbitrary scalar field $\varphi(r)$ existed. \cite{Kylie2023} Plugging in some equation for this $\varphi$ should then theoretically produce a unique $\vec{J}$ that can then be divided by the Born rule $|\vec{\Psi}(r,t)|^{2}$ for an equation of motion, velocity $\vec{v}(r,t)$. The original method did not yield dimensionally correct velocity equations [$L T^{-1}$], so the equations were then dimension fixed by multiplying be the quantities needed to get a velocity. Then reverse working to get back to a new scalar fields which had dimensions [$L^{2} T^{-1}$]. The complication therefore Pilot-Wave theory has, is that by assuming current density to have some real and physical manifestation of velocity. Then these equations that have been found possibly also hold some weight in determining the validity of Pilot-Wave Theory as there could theoretically be an infinite amount of different equations of motion a particle could have due to constant addition of more or different divergence free current densities into the continuity equation. 

\section{Preliminary}
\label{Prelim}
\subsection{Pilot-Wave Mechanics}
\label{CPWM}
It is taken that, the probability density ($\rho$) is equal to the wave function ($\vec{\Psi}$) squared, also known as the Born rule. Where ($t$) is time and ($r$) is the position.
\begin{equation}
    \rho(r) = |\vec{\Psi}(r,t)|^{2}
\end{equation}
Where, 
\begin{equation}
    |\vec{\Psi}(r,t)|^{2} = \vec{\Psi}(r,t) \cdot \vec{\Psi}^{*}(r,t)
\end{equation}
it can then be shown that by squaring the wave-function in the Schrodinger equation the same way, you can derive the equation for $\vec{J}$, where ($\vec{J}$) is the probability current density. \cite{griffiths2017introduction} \cite{current8}
\begin{equation}
\label{eqn:currentdensity}
    |\vec{\Psi}(r,t)|^{2}\cdot \vec{v}(r,t) = \frac{\hbar}{2mi}\Big(\frac{\partial\vec{\Psi}^{*}(r,t)}{\partial r}\vec{\Psi}(r,t) - \frac{d\vec{\Psi}(r,t)}{\partial r}\vec{\Psi}^{*}(r,t)\Big) = \vec{J}(r,t)
\end{equation}
where ($\hbar$) is the reduced Planck's constant, ($m$) is mass and ($\vec{v}$) is velocity. By re-arranging we then get, 
\begin{equation}
\label{eqn:velocity}
    \frac{\vec{J}(r,t)}{|\vec{\Psi}(r,t)|^{2}} = \vec{v}(r,t)
\end{equation}
the continuity equation would be as follows. 
\begin{equation}
\label{eqn:continuityEQ}
    \frac{d|\vec{\Psi}(r,t)|^{2}}{dt} + \vec{\nabla}\cdot\vec{J} = 0
\end{equation}
\newpage
\subsection{Group Velocity}
The main work done by David Bohm made use of Hamilton-Jacobi mechanics utilizing the wave function in terms of phase space $S$. Then deriving particle velocity from that. The phase space version of the wave function would take the form. 
\begin{equation}
    \vec{\Psi}(r,t) = \vec{\Psi}(r)\cdot\exp\Big(\frac{i\vec{S}}{\hbar}\Big)
\end{equation}
Substituting this equation into equation (\ref{eqn:currentdensity}) a velocity equation can be derived. 
\begin{equation}
    \vec{v}(r,t) = \frac{\vec{\nabla}S}{m}
\end{equation}
\label{eqn:CanonVel}
The implications of this essentially mean that a wave-function in a non-entangled state has no particle velocity. This can be shown via the ground state time-dependent quantum harmonic oscillator equation. \cite{holland1995quantum}
\begin{equation}
    \vec{\Psi}(r,t) = \Big(\frac{m\omega}{\pi\hbar}\Big)^{\frac{1}{4}} \cdot \exp\Big(-\frac{m\omega}{2\hbar}r^{2}\Big) \cdot \exp\Big(-i\Big(n+\frac{1}{2}\Big)\omega t\Big)
\end{equation}
\begin{equation}
    S = \Big(n + \frac{1}{2}\Big)wt
\end{equation} 
\[
\vec{\nabla}S = \frac{d}{dr} \Big(n + \frac{1}{2}\Big)wt = 0
\]
\[
\therefore \vec{v}(r,t) = 0
\]
\subsection{Simplifications via Substitutions}
Because the equations for a quantum harmonic oscillator are going to undergo vector calculus operations, a few sections are going to be substituted in order for workflow to be easier to follow and to condense clutter. $\vec{\Psi}(r,t)_{Q}$ is the wavefunction of the time-independent ground state quantum harmonic oscillator.

\begin{equation}
  \vec{\Psi}(r,t)_{Q} = \lambda \cdot \exp\big(-\beta r^{2}\big)  
\end{equation}
Where, 
\[
\lambda = \Big(\frac{m\omega}{\pi\hbar}\Big)^{\frac{1}{4}}
\]
and, 
\[
\beta = \frac{m\omega}{2\hbar}
\]
\newpage
\section{Method and Derivation}
\label{Method}
\subsection{Addition of a Divergence Free Term}
\label{DFT}
It can be shown that adding a divergence free $\vec{J}$ term to equation (\ref{eqn:continuityEQ}) would still satisfy the continuity equation as follows.

\begin{equation*}
    \frac{d|\vec{\Psi}(r,t)|^{2}}{dt} + \vec{\nabla}\cdot\vec{J}(r,t) + (\vec{\nabla}\cdot\vec{J} = 0)  = 0
\end{equation*}
From this point on it will denoted that ($\vec{\nabla}\cdot\vec{J}= 0$) is $\vec{J}_{free}$, knowing that group velocity is 0 means that equation (\ref{eqn:velocity}) must also be 0, 
\[
\frac{\vec{J}(r,t)}{|\vec{\Psi}(r,t)|^{2}} = 0
\]
thus, 
\begin{equation}
\label{eqn:NewVel}
    0 + \frac{\vec{J}(r,t)_{free}}{|\vec{\Psi}(r,t)|^{2}} = \vec{v}(r,t)_{new}
\end{equation}
$\vec{J}_{free}$ does not necessarily have to be the same as $\vec{J}$ and it will actually be shown that is the case. Because the first term is 0, it will be disregarded in all remaining derivations. \cite{Kylie2023}
\subsection{Divergence free 2D Current Density}
There are two ways to find the equation for a 2D divergence free current density. It can be derived using Cartesian coordinates first then converting to polar or by deriving straight from polar coordinates. In this report the latter will be used. \newline

\noindent It is known that the divergence of a curl is 0,
\[
\vec{\nabla}\cdot(\vec{\nabla}\times\vec{A}(r,\theta,z))=0
\]
Starting with the polar curl equation. 
\begin{equation*}
    \vec{\nabla}\times\vec{A}(r, \theta, z) = \Big(\frac{1}{r}\frac{\partial A_{z}}{\partial \theta} - \frac{\partial A_{\theta}}{\partial z}\Big)\hat{r} + \Big(\frac{\partial A_{r}}{\partial z}-\frac{\partial A_{z}}{\partial r}\Big)\hat{\theta} + \frac{1}{r}\Big(\frac{\partial (rA_{\theta})}{\partial r} - \frac{\partial A_{r}}{\partial \theta}\Big)\hat{z}
\end{equation*}
Because only 2D is of concern any $z$ components can be canceled leaving. 

\begin{equation}
\label{eqn:2dcurl}
    \vec{\nabla}\times\vec{A}(r, \theta) = \Big(\frac{1}{r}\frac{\partial A_{z}}{\partial \theta}\Big)\hat{r} + \Big(-\frac{\partial A_{z}}{dr}\Big)\hat{\theta}
\end{equation}
One single component of a vector is a scalar value, setting $A_z$ as, 
\[
A_{z} = -\varphi(r,\theta)
\]
Therefore the curl of the vector potential can become, 
\begin{equation}
\label{eqn:2dcurlinv}
  \vec{\nabla}\times\vec{A}(r, \theta) = \Big(-\frac{1}{r}\frac{\partial \varphi(r,\theta)}{\partial \theta}\Big)\hat{r} + \Big(\frac{\partial \varphi(r,\theta)}{dr}\Big)\hat{\theta}  
\end{equation}
It can been seen that equations (\ref{eqn:2dcurl}) and (\ref{eqn:2dcurlinv}) are just gradient equations for the scalar fields $\varphi$ that have been rotated by a matrix, 
\begin{equation}
    S = \begin{bmatrix} 
    0&-1 \\ 
    1&0 
    \end{bmatrix}
\end{equation}
Simplifying this gives an equation of, 
\begin{equation}
    \vec{\nabla}\times\vec{A}(r,\theta) = S\vec{\nabla}\varphi(r,\theta)
\end{equation}
Because the curl of $\vec{A}$ is divergence free, it satisfies the continuity equation (\ref{eqn:continuityEQ}) meaning it's a valid equation for $\vec{J}_{free}$, hence. \cite{2dhodge}\cite{Kylie2023}

\begin{equation}
    \vec{J}_{free} = S\vec{\nabla}\varphi(r,\theta)
\end{equation}

\subsection{Substituting Equations for Scalar Potentials}
\label{subsection:EQphi}
Because $\varphi$ is essentially ambiguous, an arbitrary starting equation can be put in place of $\varphi$. The Born rule for a 2D quantum harmonic oscillator in ground state was chosen. 
\begin{equation}
\label{eqn:QBornDist}
    |\vec{\Psi}(r,t)|^{2}_{Q} = \lambda^{2} \cdot \exp\big(-2\beta r^{2}\big) 
\end{equation}
This then gets plugged into equation (\ref{eqn:2dcurlinv}) giving, 
\begin{equation}
   \vec{J}(r,t) = \Big(-\frac{1}{r}\frac{\partial |\vec{\Psi}(r,t)|^{2}_{Q}}{\partial \theta}\Big)\hat{r} + \Big(\frac{\partial |\vec{\Psi}(r,t)|^{2}_{Q}}{dr}\Big)\hat{\theta}
\end{equation}
The first term cancels and then second term solves via chain rule to become,
\[
\Big(-\frac{1}{r}\frac{\partial |\vec{\Psi}(r,t)|^{2}_{Q}}{\partial\theta}\Big)\hat{r} \cdot \Big[\lambda^{2} \cdot \exp\big(-2\beta r^{2}\big)\Big] = 0
\] 
\begin{equation}
  \Big(\frac{\partial |\vec{\Psi}(r,t)|^{2}_{Q}}{\partial r}\Big)\hat{\theta}\cdot \Big[ \lambda^{2} \cdot \exp\big(-2\beta r^{2}\big)\Big] = \Big[\lambda^{2} \cdot 4\beta \exp\big(-2\beta r^{2}\big)\cdot r\Big] \hat{\theta}
\end{equation}
Dividing by the Born rule then should give a velocity as shown in equation (\ref{eqn:velocity}), this gives a nice reduced equation of. 
\[
\vec{v}(r)_{new} = -4\beta r\hat{\theta}
\]
Substituting the equation for $\beta$ gives,
\begin{equation}
\label{eqn:VelocityInvalid}
    \vec{v}(r)_{new} = -\frac{2m\omega r}{\hbar}\hat{\theta}
\end{equation}
This derivation has already been done previously as stated earlier, and the method of getting to this point has mostly stayed the same except for a few deviations. The derivation of the 2D polar coordinates for $\vec{J}_{free}$ was done much more efficiently and using substitutions made derivations easier. The first part of the equation can be ignored as shown here (\ref{eqn:NewVel}) as we know that the original current density associated velocity is 0. \cite{Kylie2023}

\subsection{Invalidity of New Equations}
It should be noted that equation (\ref{eqn:VelocityInvalid}) is actually invalid as when dimensional analysis is performed on the equation, the dimensions come out as. 
\[
\Big[\frac{M^{1}L^{1}T^{-1}}{M^{1}L^{2}T^{-1}}\Big] = [L^{-1}]
\]
This is not equivalent to velocity dimensions of $L^{1}T^{-1}$ This problem can be solved. Instead of looking for ambiguity in the equations of motion, we can look for ambiguity in the scalar fields responsible for that motion instead. Taking equation (\ref{eqn:VelocityInvalid}) and multiplying by a length squared and dividing by a time, gives the correct dimensions for velocity, leading to the equation.
\begin{equation}
\label{eqn:VelocityCorrected}
    \vec{v}(r,t)_{c} = -\frac{2m\omega r^{3}}{\hbar t}\hat{\theta}
\end{equation}
This essentially fixes the error made in reference \cite{Kylie2023} where $\vec{v}(r,t)_{c}$ is corrected velocity.

\subsection{Deriving Scalar Fields via Reverse Operations}
\subsubsection{Using a Standard Velocity}
Taking the standard equation of velocity, restricting to the $\hat{\theta}$ direction such that follows as shown in (\ref{subsection:EQphi}). The sign only determines an anti-clockwise or clockwise velocity and so either solution for the scalar field is valid, if we assume velocity to be negative we get a positive scalar field which is easier to visualize later on. Where $\vec{v}(r,t)_{s}$ is the standard velocity, 
\begin{equation}
\label{eqn:ThetaVel}
    \vec{v}(r,t)_{s} = -\frac{r}{t}\hat{\theta}
\end{equation}
multiplying by the born distribution as shown in equation (\ref{eqn:currentdensity}) where $\vec{J}(r, t)_{s}$ is the standard probability current.
\begin{equation}
    \vec{J}(r, t)_{s} = \Big[-\lambda^{2}\cdot \exp\big(2\beta r^{2}\big)\cdot \frac{r}{t}\Big]\hat{\theta}
\end{equation}
After substituting equation (\ref{eqn:2dcurlinv}) into the previous equation.
\[
\Big(\frac{\partial \varphi(r,t)_{s}}{dr}\Big)\hat{\theta} = -\lambda^{2} \cdot \exp\big(-2\beta r^{2}\big)\cdot \frac{r}{t}
\]
Which solves for, $\varphi(r,t)_{s}$ which is the standard scalar field. The $\hat{\theta}$ can be canceled from both sides.
\begin{equation*}
   \Big(\varphi(r,t)_{s}\Big) \hat{\theta}= \Big(- \lambda^{2}\int \exp\big(-2\beta r^{2}\big)\cdot \frac{r}{t}dr\Big)\;\hat{\theta}
\end{equation*}

\begin{equation*}
    \varphi(r,t)_{s} = \lambda^{2} \cdot \exp\big(-2\beta r^{2}\big)\cdot\frac{1}{4\beta t}
\end{equation*}
\begin{equation}
\label{eqn:standardPhi}
    \varphi(r,t)_{s} = \Big(\frac{m\omega}{\pi\hbar}\Big)^{\frac{1}{2}} \cdot \exp\Big(-\frac{m\omega}{\hbar}r^{2}\Big)\cdot\frac{\hbar}{2m\omega t}
\end{equation}
With dimensions; 
\begin{equation*}
    [L^{2}T^{-1}]
\end{equation*}
This method shows how one can go from a velocity to a scalar field.

\subsubsection{Using Dimension fixed Velocity}
Using the dimension fixed velocity (\ref{eqn:VelocityCorrected}) where $\beta$ is used in place of the constants, 
\[
\frac{4\beta r^{3}}{t}
\]
Where, $\vec{J}(r, t)_{c}$ and $\varphi(r,\theta)_{c}$ is the corrected probability current and the scalar field associated with equation (\ref{eqn:VelocityCorrected}).
\begin{equation}
   \vec{J}(r, t)_{c} = \Big[-\lambda^{2} \cdot \exp\big(-2\beta r^{2}\big)\cdot \frac{4\beta r^{3}}{t}\Big]\hat{\theta}
\end{equation}
Then integrate with respect to $dr$ to inverse the gradient of $\varphi$, as shown here,
\[
\Big(\frac{\partial \varphi(r,\theta)}{dr}\Big)\hat{\theta} = -\lambda^{2} \cdot \exp\big(-2\beta r^{2}\big)\cdot \frac{4\beta  r^{3}}{t}\hat{\theta}
\]
The theta unit vector can be canceled and then re-arranged for the integral as before, 
\[
\varphi(r,t )_{c} =-\lambda^{2}\cdot\int \exp\big(-2\beta r^{2}\big)\cdot \frac{4\beta  r^{3}}{t} dr
\]
Solving for. 
\[
\varphi(r,t)_{c} = \Big(\lambda^{2}\cdot\exp\big(-2\beta r^{2}\big)\cdot \frac{r^{2}}{t}\Big) +\Big(\lambda^{2}\cdot\exp\big(-2\beta r^{2}\big)\cdot\frac{1}{2\beta t}\Big)
\]
\begin{equation}
\label{eqn:DualField}
   \varphi(r,t)_{c} = \Big[\Big(\frac{m\omega}{\pi\hbar}\Big)^{\frac{1}{2}}\cdot\exp\Big(-\frac{m\omega r^{2}}{\hbar}\Big) \cdot \frac{r^{2}}{t}\Big] +\Big[\Big(\frac{m\omega}{\pi\hbar}\Big)^{\frac{1}{2}}\cdot\exp\Big(-\frac{m\omega r^{2}}{\hbar}\Big)\cdot\frac{\hbar}{m\omega t}\Big]
\end{equation}
With dimensions; 
\begin{equation*}
    [L^{2}T^{-1}] + [L^{2}T^{-1}]
\end{equation*}
This method shows two examples of completely different equations of motion that solve for two unique scalar field equations $\varphi$. It should be noted the previous equation is important because it holds two separate scalar fields with the same dimensions. Therefore it is valid to split this field into two separate fields and find a velocity for either side. This equation will be referred to as the dual scalar field.

\newpage

\subsubsection{Splitting the Dual Scalar Field}
It can also be seen that (\ref{eqn:standardPhi}) is the same as equation (\ref{eqn:DualField}) without the multiplication of 2. This means that the velocity it results in should also be the same as our standard velocity doubled. This subsection will show this to be the case, starting with the right-hand side equation. Such that $\varphi(r,t)_{RC}$ is the right hand-side of the corrected scalar field, 
\begin{equation}
\label{eqn:RScalarField}
    \varphi(r,t)_{RC} = \lambda^{2}\cdot\exp\big(-2\beta r^{2}\big)\cdot\frac{1}{2\beta t} 
\end{equation}
We can see this equation is identical to (\ref{eqn:standardPhi}) but multiplied by 2, Plugging into equation (\ref{eqn:2dcurlinv}) and solving gives a $\vec{J}_{free}$
\[
\Big(\frac{\partial \varphi(r,t)_{RC}}{\partial r}\Big)\hat{\theta}\cdot \Big[\lambda^{2}\cdot\exp\big(-2\beta r^{2}\big)\cdot\frac{1}{2\beta t}\Big]
\]
\begin{equation}
    \vec{J}(r, t)_{(RC)} = \Big[-\lambda^{2}\cdot 2\exp\big(-2\beta r^{2}\big)\cdot\frac{r}{t}\Big]\hat{\theta}
\end{equation}
dividing by the Born rule gives. 
\begin{equation}
\label{eqn:VelRFree}
    \vec{v}(r,t)_{RC} = -\frac{2r}{t}\hat{\theta}
\end{equation}
This is obviously just the same as equation (\ref{eqn:ThetaVel}) multiplied by 2 as expected, so it can't be considered to be unique. Moving on to the left hand-side, 

\begin{equation}
\label{eqn:LScalarField}
    \varphi(r,t)_{LC} = \lambda^{2}\cdot\exp\big(-2\beta r^{2}\big)\cdot \frac{r^{2}}{t}
\end{equation}
as before, applying equation (\ref{eqn:2dcurlinv}), 
\[
\Big(\frac{\partial \varphi(r,t)_{LC}}{\partial r}\Big)\hat{\theta} \cdot \Big[\lambda^{2}\cdot\exp\big(-2\beta r^{2}\big)\cdot \frac{r^{2}}{t}\Big]
\]
\begin{equation}
    \vec{J}(r,t)_{LC} = \lambda^{2} \cdot \Big(\frac{-4\beta \exp\big(-2\beta r^{2}\big)\cdot r^{3} + 2\exp\big(-2\beta r^{2}\big) \cdot r}{t}\Big)
\end{equation}
Dividing by the Born rule.
\[
    \Big(\frac{2r\hbar}{\hbar t}-\frac{4\beta r^{3}}{t}\Big)\hat{\theta}
\]
\begin{equation}
\label{eqn:VelLFree}
   \vec{v}(r,t)_{LC} = \Big(\frac{2r}{t}-\frac{2m\omega r^{3}}{ht}\Big)\hat{\theta}
\end{equation}
This equation turns out to be summation of (\ref{eqn:VelRFree}) and the corrected velocity (\ref{eqn:VelocityCorrected}) which is ends up as a 3rd unique equation of motion, taking a dual form similar to the dual scalar field. As such combining (\ref{eqn:VelRFree}) and (\ref{eqn:VelLFree}) then reduces back to the corrected velocity equation. (\ref{eqn:VelocityCorrected}).
\[
-\frac{2m\omega r^{3}}{ht}\hat{\theta}
\]
Which shows that this method is reversible.

\section{Discussion and Plots}

\subsection{Features of Pilot Wave Theory}
Mathematically speaking the equations in this report can be derived and applied to all interpretations due to all of them using the Born rule, and therefore a continuity equation can always be formulated. There is a particular reason why Pilot-Wave theory is being verified in this report specifically. it's the only interpretation that assumes that physical particles exist with definite positions, and makes use of Hamilton-Jacobi mechanics to get particle trajectories. Unlike other interpretations, where the use of phase space and velocity is only used in the context of visualization of the evolution of probability current over time. Because most interpretations don't assume the particle existence until observed, any calculated alternate velocities are not physical or real in a meaningful way and disappear upon wavefunction collapse. Using an isolated system like the one used in this report, the observation of a particle gives you its position, then the equations derived in section (\ref{Method}) would then tell you its velocity. Because wavefunction doesn't collapse for Pilot-Wave theory, these velocity equations are always valid solutions. 

\subsection{Duality and Radial Velocities}
One of the main points that should first be discussed, is the duality nature that exists for either the equation of motion or scalar potential. Specifically for the equations that were used in this scenario. The standard velocity gives us the standard scalar field equation (\ref{eqn:standardPhi}), that is also present in the dual scalar field. Similarly the left-hand side velocity was a dual velocity equation $\vec{v}(r,t)_{LC}$ (\ref{eqn:VelLFree}), that contained the standard velocity (\ref{eqn:ThetaVel}). This would suggest that any dimension fixed fields may be linked to the standard fields fundamentally. This could simply be due to how the dimension fixed velocity was derived, by multiplying by the variables needed. So although multiple equations of motion and scalar fields found were unique. It appears that the standard cases are the only ones that can exist by themselves without an accompanying field. It does seem that the idea of using the born rule to begin with as a scalar field had some merit, but this can be expanded by essentially putting any equation in its place and seeing what comes out. Then if needs be, dimension fixing and working backwards to see if the same pattern emerges. it must be stated the the scalar field does follow dimensions [$L^{2} T^{-1}$] so any equation that is put in there will have to adhere to that due to the nature of the gradient operator. More research will have to be done to see if this always works and if the scalar and velocity fields produced are also linked to their standard cases as mentioned previously. The equations used have velocities in the theta ($\hat{\theta}$) direction predicting an orbital nature There should theoretically be velocity equations that have radial directions ($\hat r$), for scalar fields that have a angular variable $\theta$ within them, as can be seen in equation (\ref{eqn:2dcurlinv}) 

\subsection{Multi-State Entanglement Velocities and Quantum Non-Equilibrium}
In this report only the ground state of the quantum harmonic oscillator was focused on, assuming no entanglement. When entanglement between states is accounted for particle motion will manifest naturally. Which can be derived via the equations in section (\ref{Prelim}) without the need for divergence free probability current densities. One could argue that until observed the particle is entangled with all states so must have a velocity. This could be explored in further detail in future research, including the analysis of divergence free probability current densities within an entangled system. Pilot-wave theory allows for quantum non-equilibrium, where the probability density is not equal to the born rule, $\rho \not= |\vec{\Psi}(r,t)|^{2}$. Although these non equilibrium systems have been shown to relax back to equilibrium after a certain amount of time \cite{underwood2019signaturesrelicquantumnonequilibrium}. There are scenarios where the system does not relax back to equilibrium, where stability is possible in plane and Gaussian waves \cite{Colin_2010}. There is also the possibility that relic non-equilibrium states could still exist from the rapid inflationary period in the early universe \cite{Underwood_2015}. Recent papers have shown non-equilibrium could be preserved either permanently and for long periods in coupled systems of 1D harmonic oscillators \cite{Lustosa_2023}. If such exotic states can possibly exist then it is fair to assume that the addition of a divergence free current density and its associated ambiguous equations of motion is not a radical inclusion to the theory. Mathematical predictions should be taken with skepticism though, as experimental evidence has not yet shown the validity of these predictions.

\subsection{Particle Velocity Behavior}
It can be seen via all the velocity equations obtained that the further out a particle is measured to be, the faster it goes. It seems counter intuitive as orbital nature would have slower velocities the further out the particle is from the center. This paradox is solved knowing the system is a harmonic oscillator, as the time period must remain constant. This means larger distances would have to be covered by greater velocities \cite{Kylie2023}. Leading to some extreme velocities at the very edges of this system. Given a relatively large radii above 2nm (nanometers) velocities would exceed the speed of light, this may be corrected with the use of special relativity using a relativistic harmonic oscillator. Regardless of that fact the probability that a particle is even observed at these edges is extremely small. Velocity does increase much more rapidly in the dimension fixed velocity equation, therefore relativistic velocities do appear within reasonably small distances. For the standard velocity equation the increase in velocity is much slower. This could give some merit towards the standard velocity equation being a more fundamental solution. As previously mentioned there are many different equations that could be used for the scalar field and infinitely many additions of $\vec{J}_{free}$ can be applied to the continuity equation.

\subsection{Scalar Field Plots}
In order to plot the 3D graphs, chosen values of parameters had to be plugged in first, energy, distance and mass, this was also done in reference \cite{Kylie2023}. A chosen range of -1 nanometer (nm) to 1nm was used for distance and an energy ($E$) of 1 electron volt (eV), it was decided that an electron would be used for the modeled particle which has a known mass. Because the system is a quantum harmonic oscillator the energy can be used to calculate the angular velocity ($\omega$) and time period ($t$)  using,
\begin{equation}
    \omega = \frac{2E}{\hbar}
\end{equation}
and,
\begin{equation}
    t = \frac{\pi\hbar}{E}
\end{equation}
 
\newpage

\subsubsection{Standard Scalar Field Plot}
Plotting the scalar field $\varphi(r,t)_{s}$ (\ref{eqn:standardPhi}) associated with the standard velocity equation (\ref{eqn:ThetaVel}) 
\begin{figure}[H]
    \centering
    \includegraphics[scale=0.4]{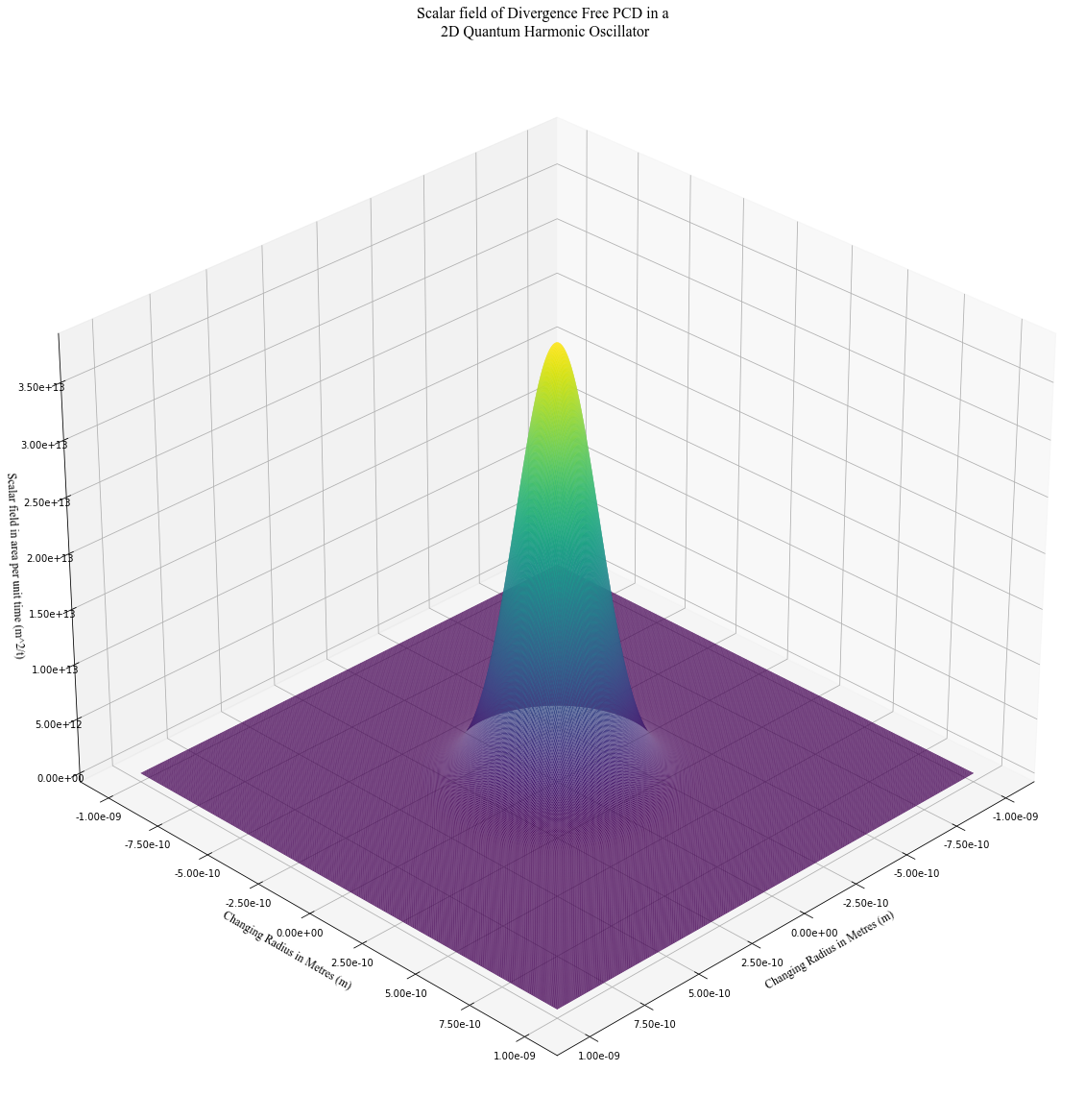}
    \caption{Shows the 3D plot for the standard scalar field $\varphi(r,t)_{s}$ associated with a 2D quantum harmonic oscillator, bound in a ground state energy level in the range of -1nm to 1nm and 1eV of energy using equation (\ref{eqn:standardPhi})}
    \label{img:SSF}
\end{figure}

\newpage 

\subsubsection{Right-Hand Side Dual Scalar Field Plot}
Plotting the scalar field $\varphi(r,t)_{RC}$ (\ref{eqn:RScalarField}) associated with the dimension fixed velocity equation (\ref{eqn:VelocityCorrected}) 
\begin{figure}[H]
    \centering
    \includegraphics[scale=0.4]{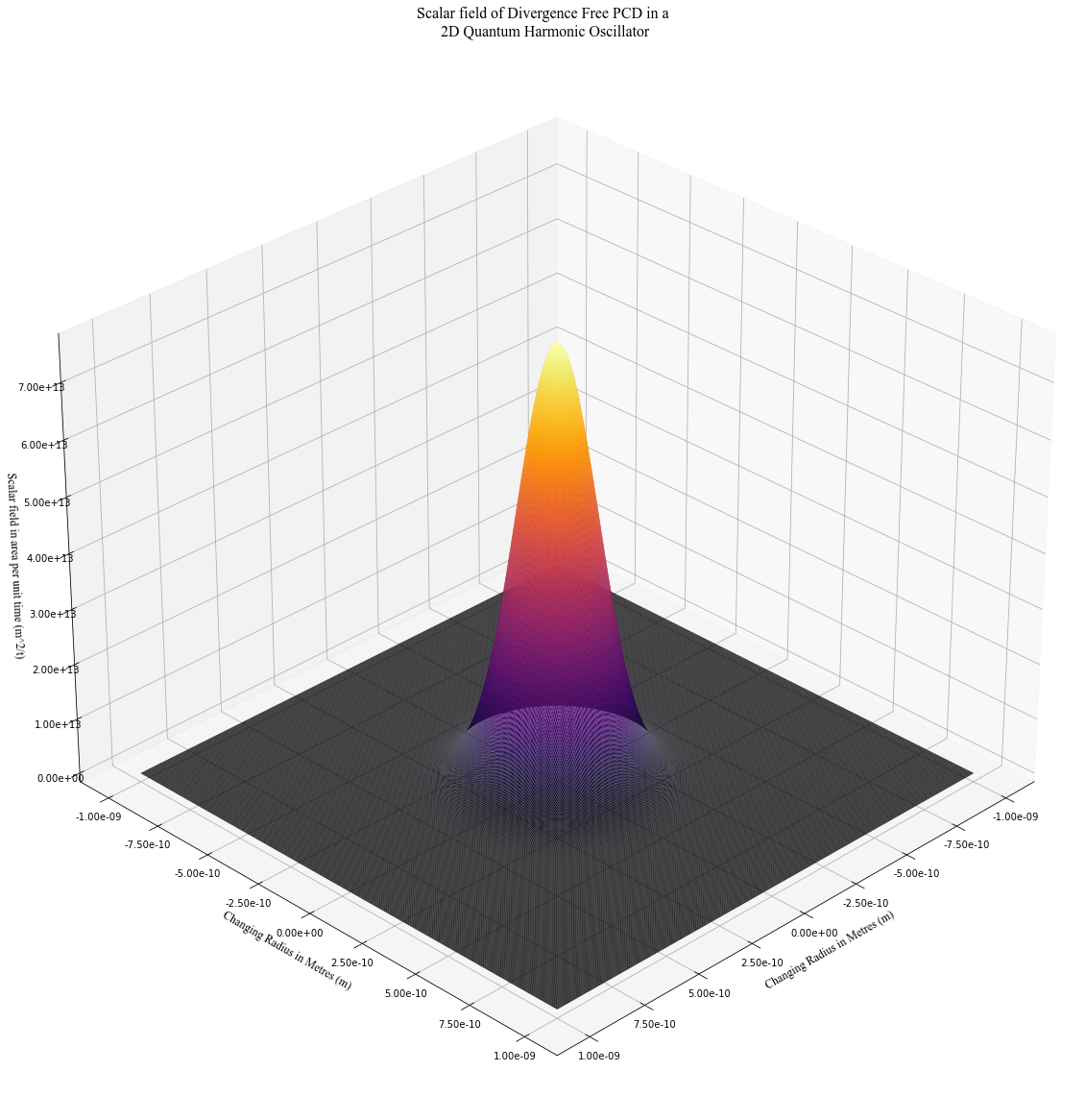}
    \caption{Shows the 3D plot for the right-hand side of the dual scalar field $\varphi(r,t)_{RC}$ associated with a 2D quantum harmonic oscillator, bound in a ground state energy level in the range of -1nm to 1nm and 1eV of energy using equation (\ref{eqn:RScalarField})}
    \label{img:RHSD}
\end{figure}

\newpage

\subsubsection{Left-Hand Side Dual Scalar Field Plot}
Plotting the scalar field $\varphi(r,t)_{LC}$ (\ref{eqn:LScalarField}) associated with the dimension fixed velocity equation. (\ref{eqn:VelocityCorrected}) 
\begin{figure}[H]
    \centering
    \includegraphics[scale=0.4]{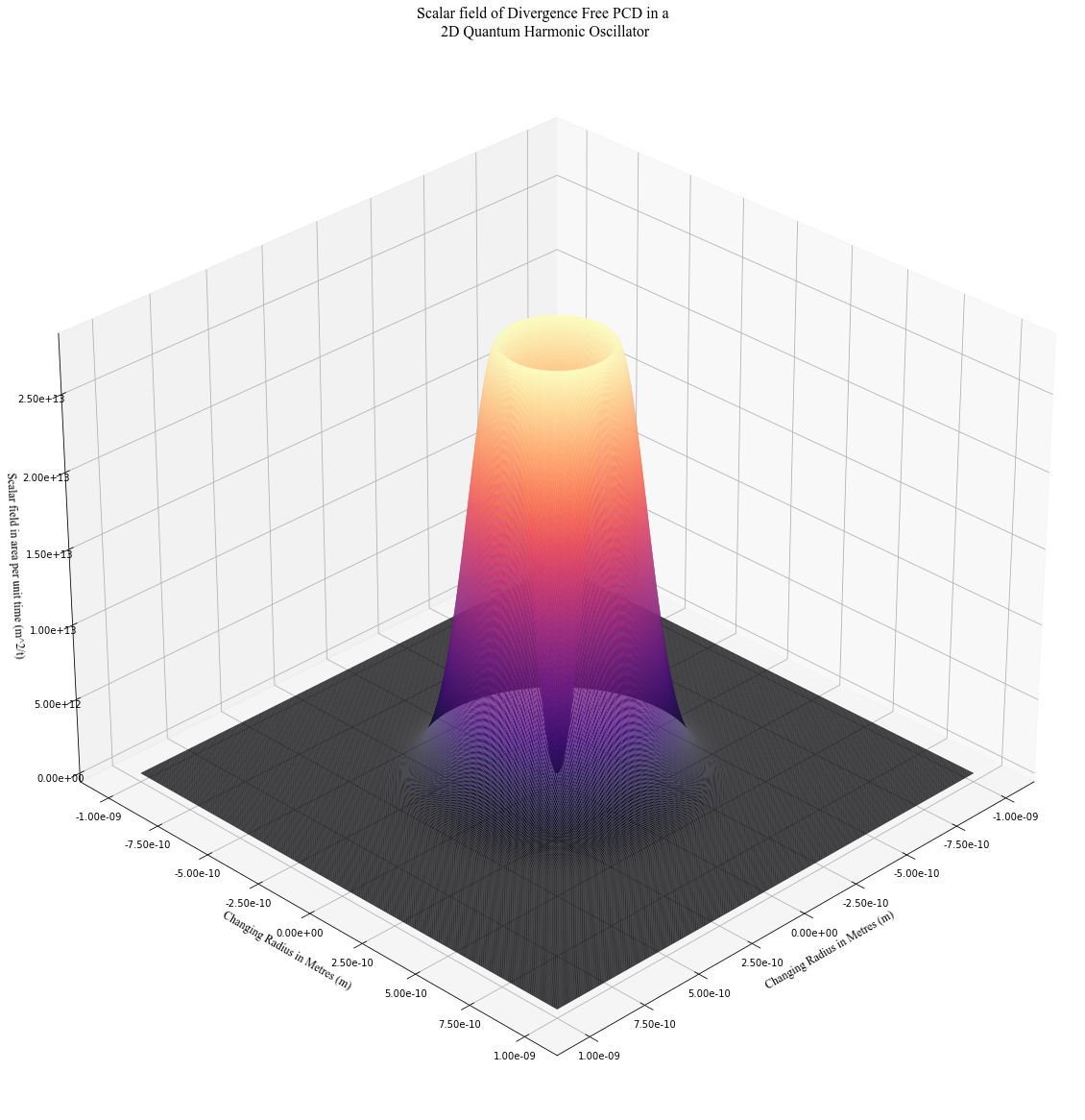}
    \caption{Shows the 3D plot for the left-hand side of the dual scalar field $\varphi(r,t)_{LC}$ associated with a 2D quantum harmonic oscillator, bound in a ground state energy level in the range of -1nm to 1nm and 1eV of energy using equation (\ref{eqn:LScalarField})}
    \label{img:LHSD}
\end{figure}

\newpage

\subsubsection{Dual Scalar Field Plot}
Plotting the scalar field $\varphi(r,t)_{C}$ (\ref{eqn:DualField}) associated with the dimension fixed velocity equation (\ref{eqn:VelocityCorrected}) 
\begin{figure}[H]
    \centering
    \includegraphics[scale=0.4]{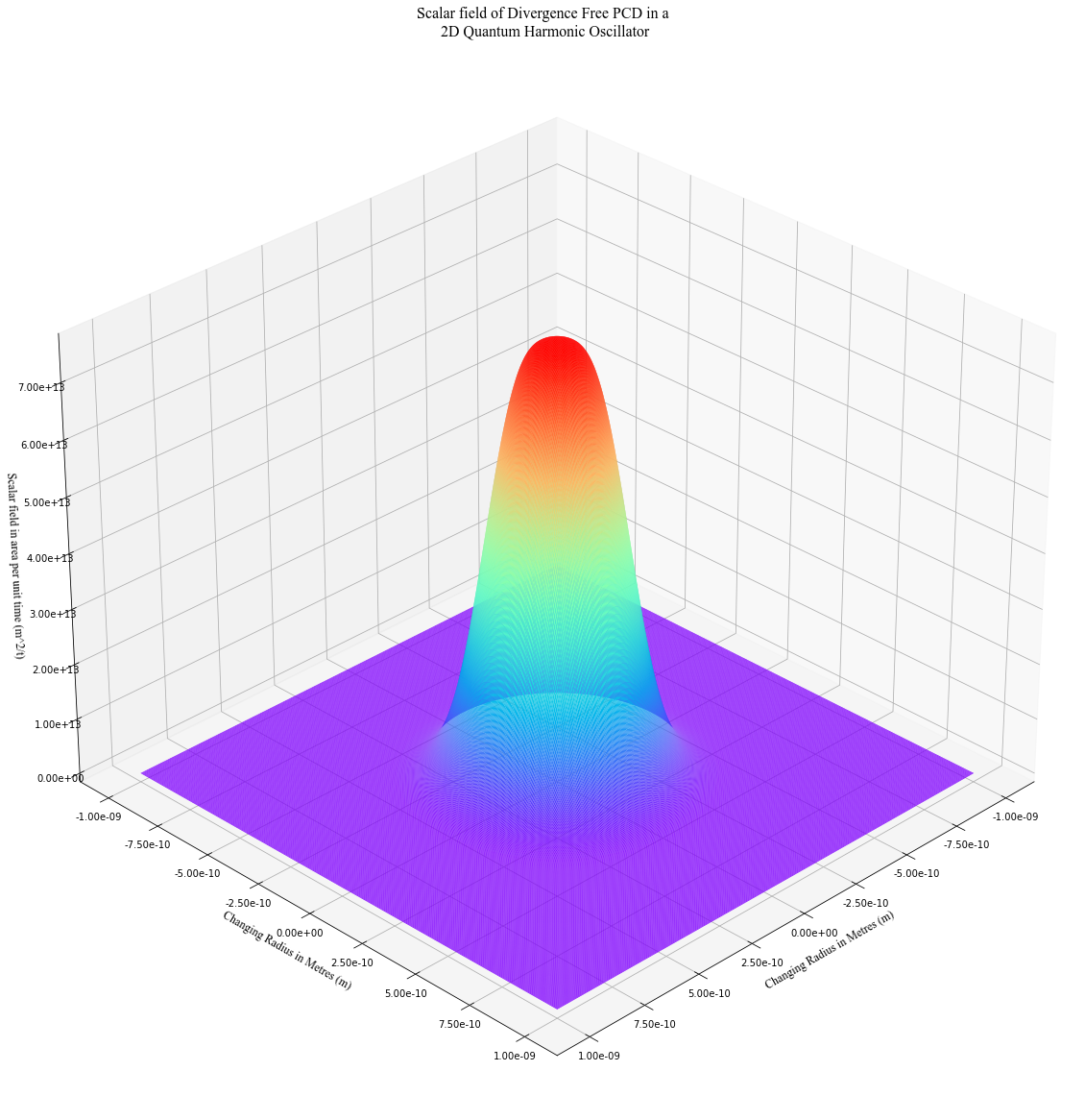}
    \caption{Shows the 3D plot for the dual scalar field $\varphi(r,t)_{C}$ associated with a 2D quantum harmonic oscillator, bound in a ground state energy level in the range of -1nm to 1nm and 1eV of energy using equation (\ref{eqn:DualField})}
    \label{img:DSFP}
\end{figure}

\newpage

\subsection{Velocity Field Plots} 

\subsubsection{Standard Velocity plot}
Plotting the velocity field $\vec{v}(r,t)_{S}$ (\ref{eqn:ThetaVel}) associated with the standard scalar field. (\ref{eqn:standardPhi})
\begin{figure}[H]
    \centering
    \includegraphics[scale=0.4]{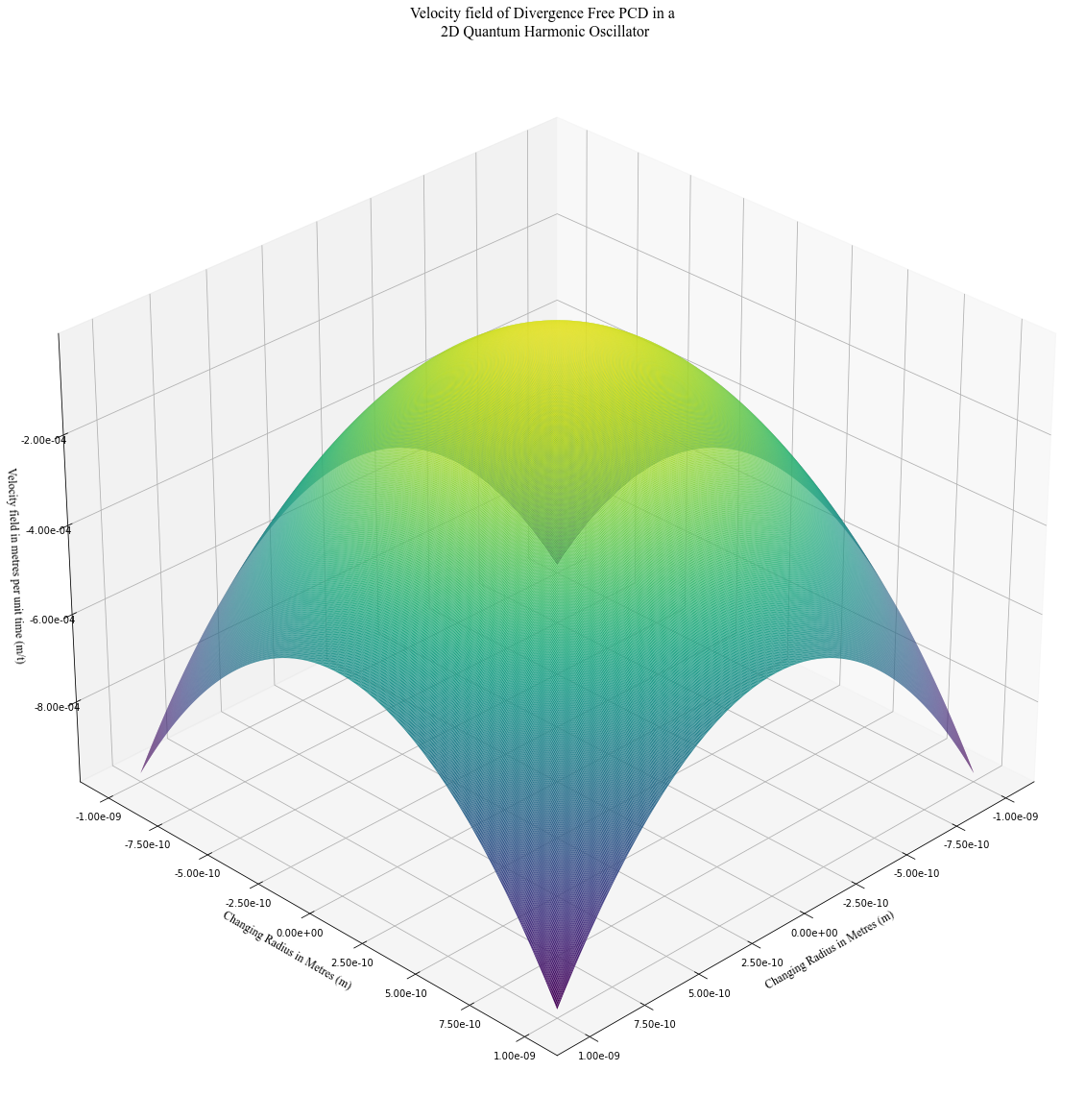}
    \caption{Shows the 3D plot for standard velocity $v(r,t)_{S}$ associated with a 2D quantum harmonic oscillator, bound in a ground state energy level in the range of -1nm to 1nm and 1eV of energy using equation (\ref{eqn:ThetaVel})}
    \label{img:SVP}
\end{figure}

\newpage

\subsubsection{Right-Hand side Dual Velocity Plot}
Plotting the velocity field $\vec{v}(r,t)_{RC}$ (\ref{eqn:VelRFree}) associated with the dual scalar field. (\ref{eqn:DualField})
\begin{figure}[H]
    \centering
    \includegraphics[scale=0.4]{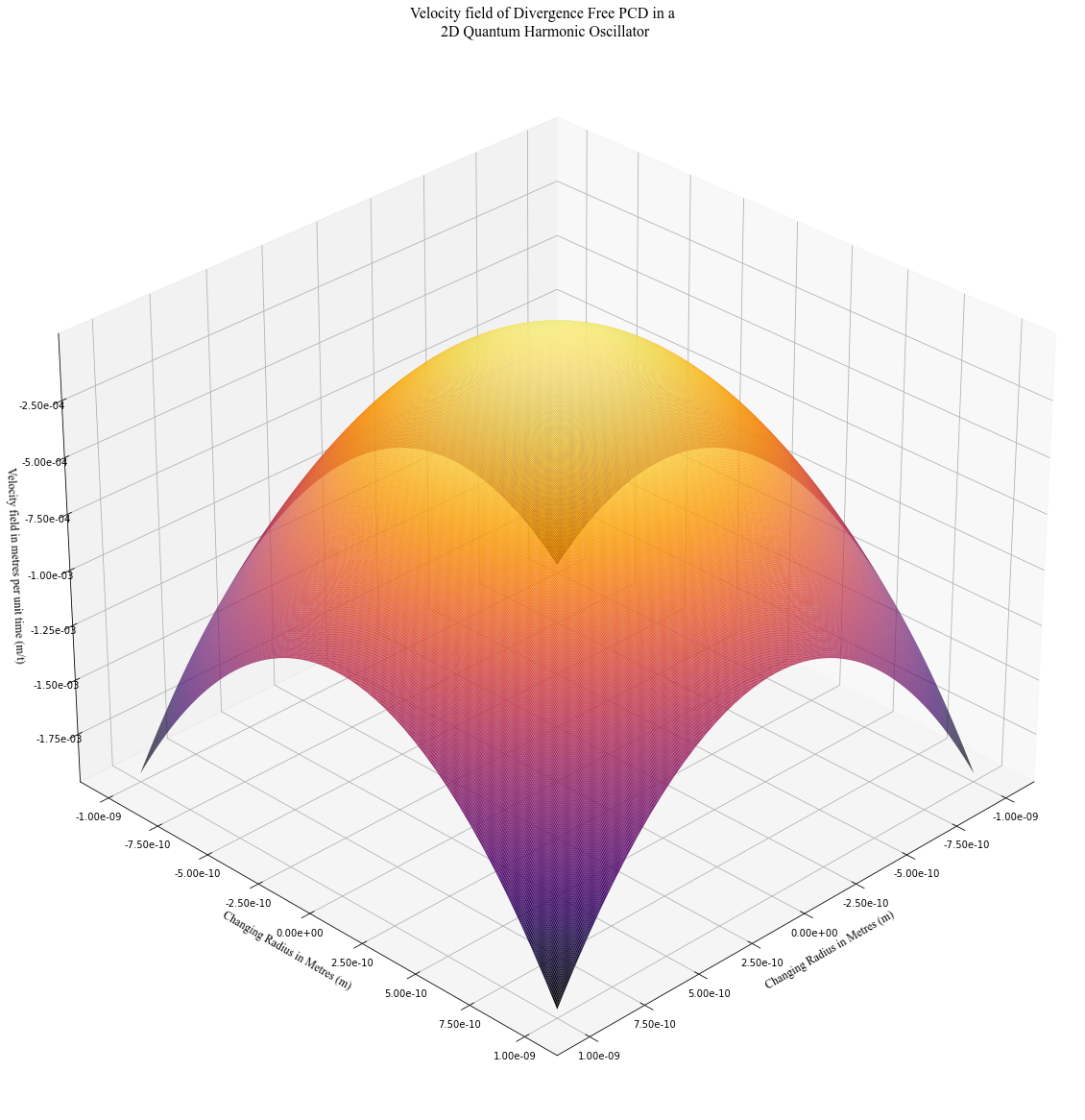}
    \caption{Shows the 3D plot for right-hand side velocity $v(r,t)_{RC}$. Derived from the dual scalar field, associated with a 2D quantum harmonic oscillator, bound in a ground state energy level in the range of -1nm to 1nm and 1eV of energy using equation (\ref{eqn:VelRFree})}
    \label{img:VFR}
\end{figure}

\newpage

\subsubsection{Left-Hand Side Dual Velocity Plot}
Plotting the velocity field $\vec{v}(r,t)_{LC}$ (\ref{eqn:VelLFree}) associated with the dual scalar field. (\ref{eqn:DualField})
\begin{figure}[H]
    \centering
    \includegraphics[scale=0.4]{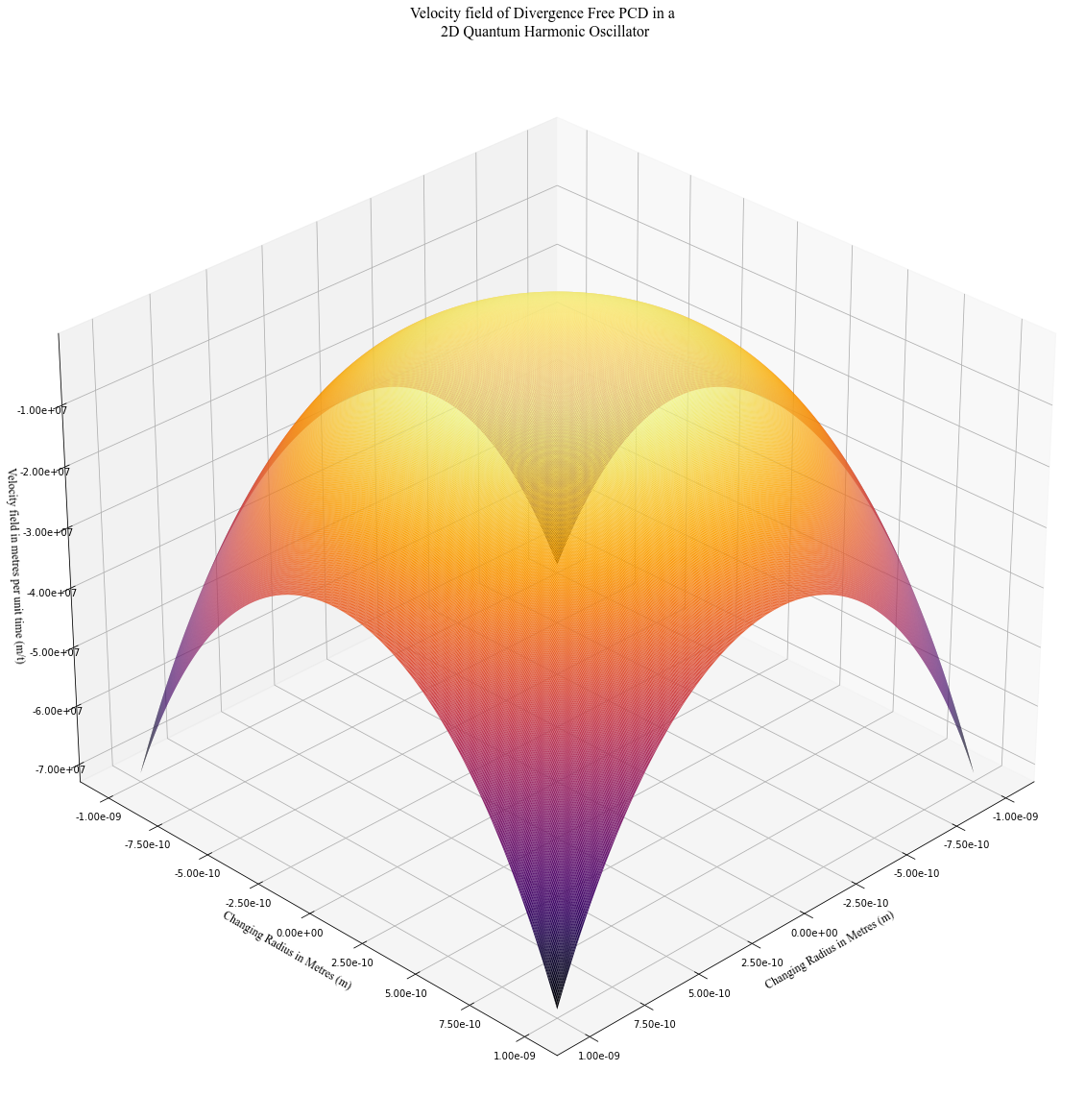}
    \caption{Shows the 3D plot for left-hand side velocity $v(r,t)_{LC}$. Derived from the dual scalar field, associated with a 2D quantum harmonic oscillator, bound in a ground state energy level in the range of -1nm to 1nm and 1eV of energy using equation (\ref{eqn:VelLFree})}
    \label{img:LHSV}
\end{figure}

\newpage

\subsubsection{Dimension Fixed Velocity Plot}
Plotting the velocity field $\vec{v}(r,t)_{C}$ (\ref{eqn:VelocityCorrected}) associated with the dual scalar field. (\ref{eqn:DualField})
\begin{figure}[H]
    \centering
    \includegraphics[scale=0.4]{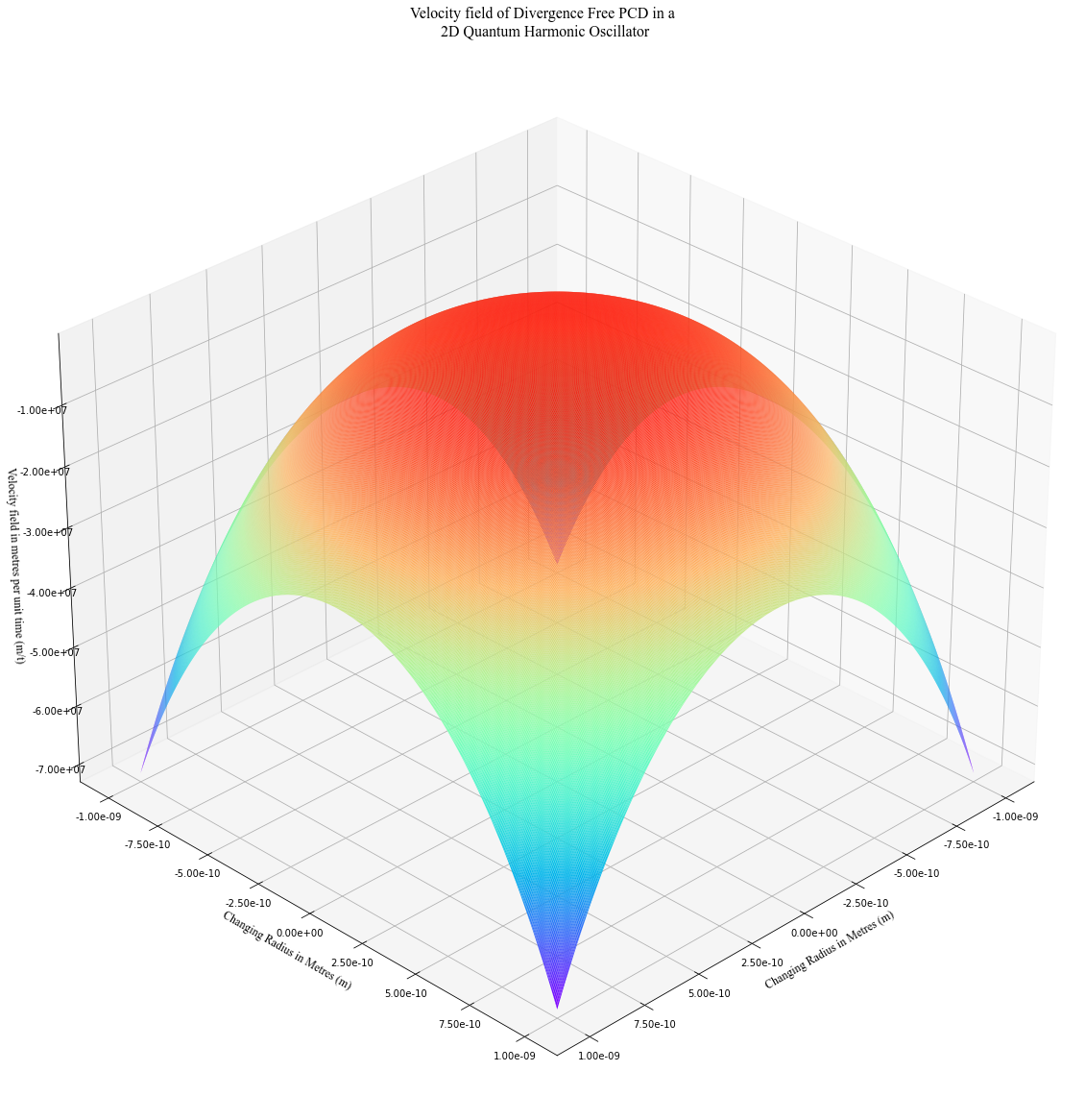}
    \caption{Shows the 3D plot for dimension fixed velocity $v(r,t)_{C}$ associated with a 2D quantum harmonic oscillator, bound in a ground state energy level in the range of -1nm to 1nm and 1eV of energy using equation (\ref{eqn:VelocityCorrected})}
    \label{img:DFV}
\end{figure}

\newpage

\section{Conclusion}
This study has explored the derivation of unique equations of motion and scalar fields within the framework of De Broglie-Bohm Pilot-Wave Theory, showing the implications of divergence-free probability density currents. By deriving and analyzing three distinct scalar fields and their corresponding equations of motion for a 2D quantum harmonic oscillator, it has shown that multiple mathematically valid solutions exist, highlighting the ambiguity in possible equations of motion. Despite identifying unique scalar fields and velocities, the results indicate that so far only the standard fields can exist independently, without requiring accompanying fields. 3 scalar fields were found that were attributed to 3 unique equations of motion. A standard velocity equation (\ref{eqn:ThetaVel}) was used to derive the first scalar field equation (\ref{eqn:standardPhi}). This was the only case found to have no extra accompanying field equations that were referred to as duals. A dimension fixed equation resulting in a velocity (\ref{eqn:VelocityCorrected}) was then used to gain a new scalar field (\ref{eqn:DualField}). That was a duel field being comprised of two separate scalar fields a left-hand side (\ref{eqn:LScalarField}) and a right-hand side (\ref{eqn:RScalarField}) that had the same dimensions. These scalar fields were split and velocities were derived from both. The left-hand side produced a dual velocity field (\ref{eqn:VelLFree}) that contained both the dimension fixed velocity and a standard velocity. The right-hand side produced a standard velocity multiplied by a factor of 2 (\ref{eqn:VelRFree}) adding both (\ref{eqn:VelLFree}) and (\ref{eqn:VelRFree}) together reduced down to the original dimension fixed velocity (\ref{eqn:VelocityCorrected}) which verified the method is reversible. 
\newline
\newline
It was determined this dual nature possibly arose because of how the corrected velocity was derived and that in order to get a clearer picture. More scenarios are going to have to be constructed, such as higher dimensional systems like the hydrogen atom or 3D harmonic oscillator. Additional equations being substituted for the scalar potential $\varphi$ should also be considered, instead of the born rule, repeating the method used in this report (if needs be) for new equations of motion. A further avenue of research could be assessing whether similar features arise in quantum non-equilibrium states, as well as the potential physical interpretations of the scalar fields themselves. Continuity equations are not unique to quantum mechanics, appearing in both electrodynamics, fluid mechanics and more. Applying the same method in this report to the continuity equations in those fields could provide some insight into what the scalar fields could be or if the method used in this report itself is valid. Overall the findings, so far, in this report do agree that a particles equation of motion in Pilot-Wave Theory is potentially ambiguous but further research will have to be done for more confirmation.
\newpage
\subsection{Acknowledgments}
I would like to thank Sam McBride for proof reading the original draft and providing insight into the grammar and structure of the writing.
\section{References}
\bibliographystyle{ieeetr}
\bibliography{Bibtex}

\end{document}